\documentclass[twocolumn,aps,showpacs,pre,amsmath]{revtex4}

\usepackage{amssymb}
\usepackage{graphicx}
\usepackage{dsfont}

\begin{document}

\title{Strain-induced modulation of Dirac cones and van Hove singularities \\ in twisted graphene bilayer}

\author{V. Hung Nguyen$^{1,2}$\footnote{E-mail: hung@iop.vast.ac.vn} and P. Dollfus$^1$} \address{$^1$Institut d'Electronique Fondamentale, UMR8622, CNRS, Universit$\acute{e}$ Paris Sud, 91405 Orsay, France \\ $^2$Center for Computational Physics, Institute of Physics, Vietnam Academy of Science and Technology, P.O. Box 429 Bo Ho, 10000 Hanoi, Vietnam}

\begin{abstract}
    By means of atomistic tight-binding calculations, we investigate the effects of uniaxial strain on the electronic bandstructure of twisted graphene bilayer. We find that the bandstructure is dramatically deformed and the degeneracy of the bands is broken by strain. As a conseqence, the number of Dirac cones can double and the van Hove singularity points are separated in energy. The dependence of these effects on the strength of strain, its applied direction and the twist angle is carefully clarified. As an important result, we demonstrate that the position of van Hove singularities can be modulated by strain, suggesting the possibility of observing this phenomenon at low energy in a large range of twist angle (i.e., larger than $10^\circ$). Unfortunately, these interesting/important phenomena have not been clarified in the previous works based on the continuum approximation. While they are in good agreement with available experiments, our results provide a detailed understanding of the strain effects on the electronic properties and may motivate other investigations of electronic transport in this type of graphene lattice.
\end{abstract}

\pacs{xx.xx.xx, yy.yy.yy, zz.zz.zz}
\maketitle

Nowadays, graphene is one of the most attractive electronic materials because of its specific electronic properties, which are a consequence of its two-dimensional honeycomb lattice as summarized, e.g., in the review \cite{neto09}. It is the basis of several peculiar phenomena and promising applications of graphene materials. More interestingly, the electronic structure of graphene is relatively easy to be tune, e.g., by strain \cite{pere09a}, substrate \cite{yank12}, perpendicular electric field \cite{zhan09} etc. The formation of Van der Waals heterostructures \cite{geim13} has been also suggested as an effective route to control the electronic structure of graphene. Multilayer graphene, formed by graphene layers only, is actually one of these van der Waals structures. To modulate its electronic structure, one can rotates one graphene layer with respect to the other ones (i.e., twisted few-layer graphene) to form graphene-on-graphene moir\'{e} patterns. The twisted few-layer graphene lattices often appear in the thermal decomposition of the C-face SiC or in the copper-assisted growth using the chemical vapor deposition method, e.g., see in \cite{heer07,hass08,caro11,luic11,have12,cclu13,sant07,shal10,lais10,guli10,brih12,sant12}. Indeed, the bandstructure in twisted graphene bilayer (T-GBL) changes dramatically \cite{sant07,shal10,lais10,guli10,brih12,sant12}, compared to that of monolayer or Bernal/AA stacking bilayer systems. In addition to the existence of linear dispersion in the vicinity of K-points, saddle points emerge at the crossing of Dirac cones, yielding van Hove singularities (VHS) in the density of states at low energies, and remarkable renormalization of the Fermi velocity is observed. Moreover, several properties as, e.g., magnetic field effects, optical properties, and phonon transport in the T-GBLs have also a strong dependence on the twist angle \cite{moon12,moon13,coce13}.

\begin{figure}[!t]
\centering
 \includegraphics[width=3.4in]{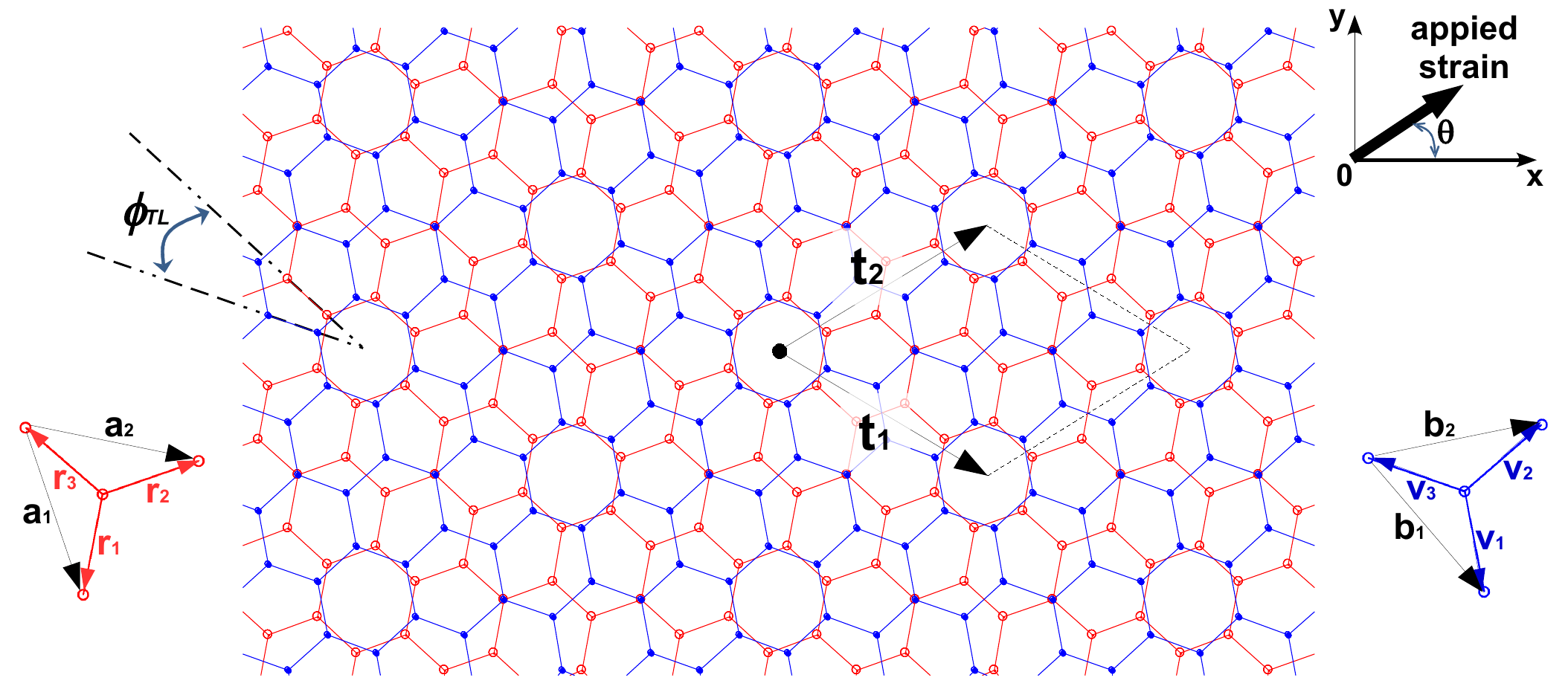}
\caption{Schematic of twisted graphene bilayer with the twist angle $\phi_{TL}$ investigated in this work. A uniaxial strain of angle $\theta$ with respect to the axis Ox is studied.}
\label{fig_sim0}
\end{figure}
In addition to its interesting electronic properties, graphene also has remarkable mechanical properties. Indeed, it is able to sustain a much larger (i.e., $> 20\%$ \cite{shar13}) strain than other semiconductors, making it a promising candidate for flexible electronics. Very recently, the techniques \cite{garz14,shio14} to generate extreme strain in graphene in a controlled and nondestructive way have been also explored. Interestingly, it has been shown that strain engineering is an efficient approach to modulating the electronic properties of graphene nanomaterials. Many interesting electrical, optical and magnetic properties induced by strain have been hence observed, e.g., see refs. \cite{pere09a,cocc10,yalu10,kuma12,baha13,hung14a,chun14,hung14b,pere10,guin10,tlow10,zhai11}. Although the bandgap of slightly strained (a few percent) 2D graphene remains zero \cite{pere09b}, the strain can be used to generate or strongly modulate transport gaps in some specific graphene channels, e.g., graphene nanoribbons with a local strain \cite{yalu10,baha13}, graphene with grain boundaries \cite{kuma12}, graphene strain junctions \cite{hung14a,chun14}, and vertical devices made of twisted graphene layers \cite{hung14b}.

\begin{figure*}[!t]
\centering
\includegraphics[width=5.4in]{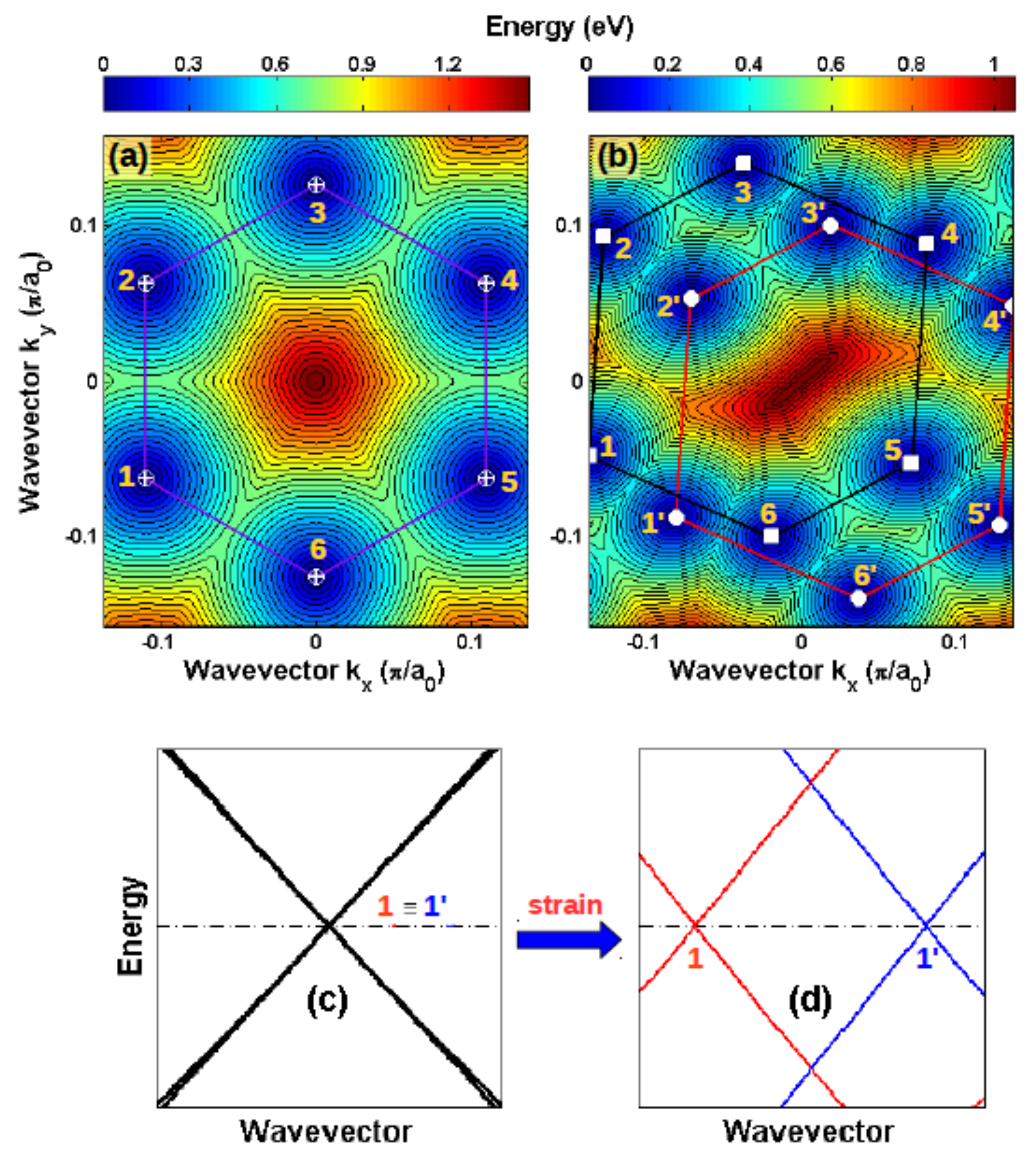}
\caption{Map of the lowest positive-energy bands of twisted graphene bilayer without (a) and with strain (b). The twist angle $\phi_{TL} = 9.43^\circ$ and strain of $\left(\sigma,\theta\right)=\left(6\%,20^\circ\right)$ are considered here. (c) and (d) show schematically the strained-induced change in the bandstructure around the \emph{K}-point.}
\label{fig_sim1}
\end{figure*}

\begin{figure*}[!t]
\centering
\includegraphics[width=6.3in]{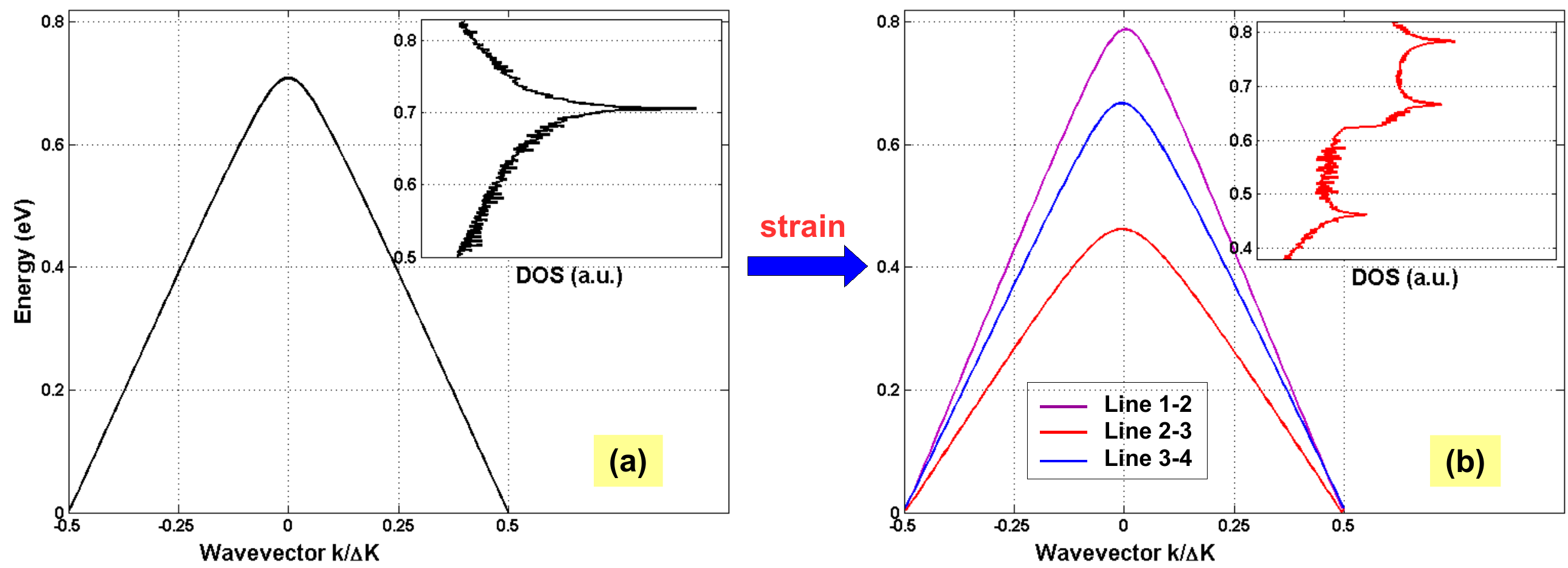}
\caption{For the same structure and strain as in Fig. 2, profile of the lowest conduction band joining the Dirac points in the (a) unstrained and (b) strained cases. The insets show the DOS as a function of energy illustrating the existence of van Hove singularities.}
\label{fig_sim2}
\end{figure*}
The bandstructure of T-GBLs has been investigated in several works (e.g., in \cite{sant07,lais10,moon12,brih12,sant12,moon13,dchu13,myan13}) using different approaches including first principles calculations, tight binding methods and continuum approximation. To the best of our knowledge, the strain effects on VHSs have been explored only in \cite{dchu13,myan13} and discussed on the basis of calculations in the continuum approximation without taking into account the details of the atomic arrangement. This approximation has been shown to give a good description of the T-GBL bandstructure at low energy and allowed for explaining the main/basic properties of VHSs observed in experiments for the slightly twisted graphene bilayer without strain. However, when strain is applied, the atomic arrangement of T-GBLs is dramatically affected and becomes very strongly anisotropic, due to the strain-induced changes in \textit{C-C} bond lengths. This can result in complicated deformation of graphene bandstructure, i.e., the strain effects are strongly dependent on the direction of applied strain and on the lattice orientation, as previously reported in refs. \cite{chun14,pere09b,kuma12,hung14b}. In particular, the strain effects on the electronic properties of two graphene sheets with different orientations should be, in principle, different. In a recent study of graphene vertical devices made of two twisted graphene layers \cite{hung14b}, we have demonstrated that the strain can lead not only to the displacement of Dirac cones from the $K-$point but also to the separation of Dirac cones of the two sheets. Interestingly, this feature results in a strain-induced finite conduction gap in these vertical channels and is strongly dependent on the strain direction. These properties should, in principle, have a strong impact on the bandstructure of T-GBLs but they have not been fully clarified yet in the previous studies based on the continuum approximation. This raises a question about the detailed effects of strain on the bandstructure of T-GBLs when employing more accurate approaches including explicitly the arrangement of atoms. In this context, our aim here is to revisit this topic, i.e., the strain effects on the bandstructure of T-GBLs, using appropriate atomistic tight-binding calculations. In this paper, we demonstrate indeed that several interesting features are still missing and/or have not been well clarified in the previous works based on the continuum approach.
\begin{figure*}[!t]
\centering
\includegraphics[width=5.2in]{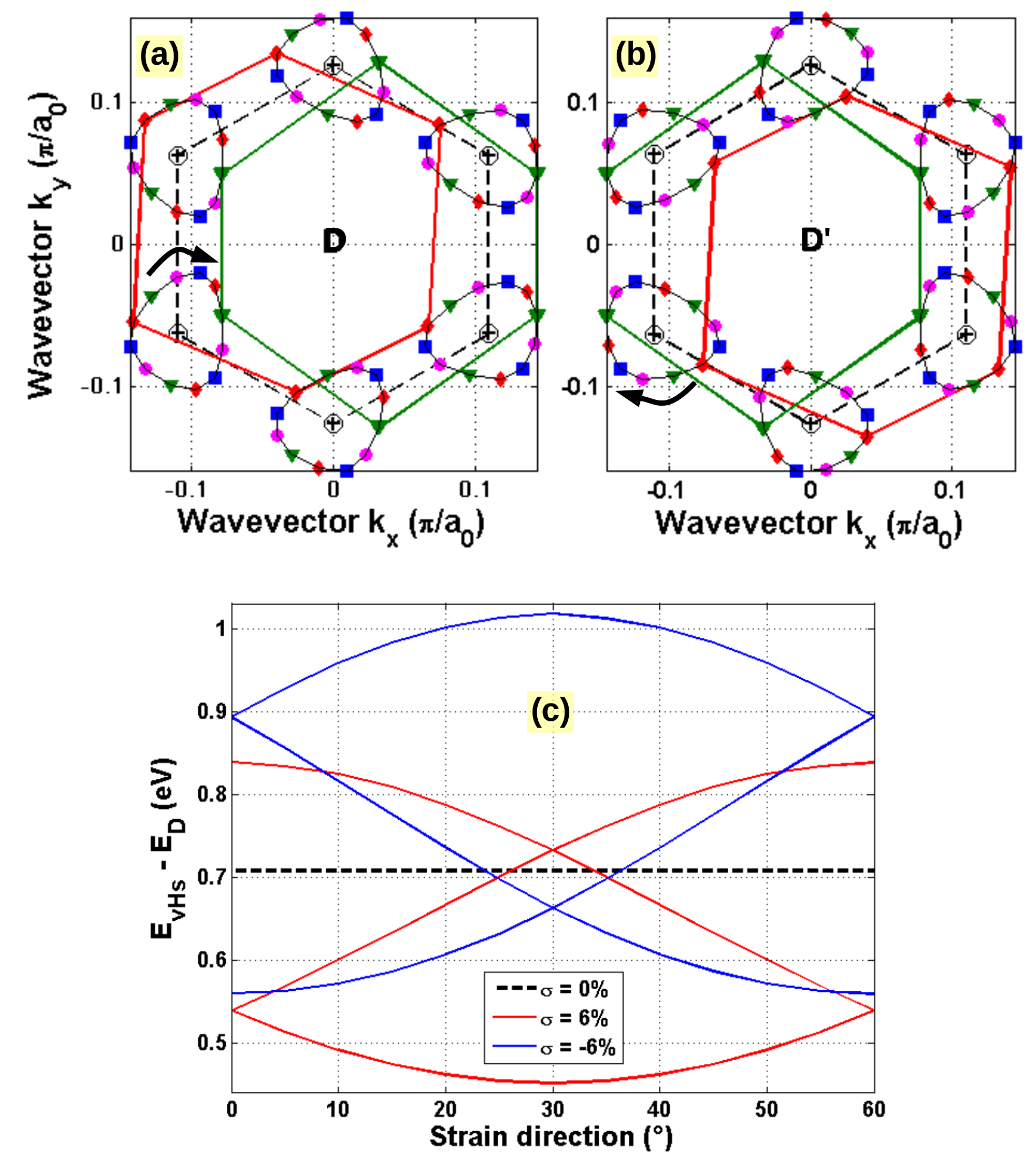}
\caption{For the twist angle $\phi_{TL} = 9.43^\circ$, diagrams showing the strained-induced displacement of Dirac cones ($D$ in (a) and $D'$ in (b), see in the text) from the $K$-points of unstrained lattice (circle-plus symbols) when changing the strain direction $\theta$ from $-90^\circ$ to $90^\circ$ with the step $\Delta\theta = 15^\circ$. The diamond red (triangle green) lines correspond to $\theta = 15^\circ$ and $\theta = 90^\circ$ ($\equiv -90^\circ$). (c) shows the energy spacing between van Hove singularity $E_{vHs}$ and neutrality $E_D$ points as a function of strain direction. The strain is $\sigma = 6\%$ in (a,b) while both tensile/compressive strains are considered in (c)}
\label{fig_sim3}
\end{figure*}

We investigate the twisted graphene lattices consisting of two parallel graphene sheets, i.e., twisted graphene bilayer (T-GBL). Originating from the AA stacked graphene bilayer, the rotation center was chosen at the hexagon center and then all lattices considered here were generated by rotating one sheet with respect to the other one by a commensurate angle $\phi_{TL}$ (see Fig. 1). This commensurate angle is determined by $\cos {\phi _{TL}} = ({n^2}/2 + 3mn + 3{m^2})/({n^2} + 3mn + 3{m^2})$ \cite{sant12}, where \emph{n} and \emph{m} are coprime positive integers. The primitive vectors of the Bravais lattice are determined as follows:
\begin{equation}
\left\{ \begin{array}{l}
{{\vec t}_1} = m{{\vec a}_1} + \left( {n + m} \right){{\vec a}_2}\\
{{\vec t}_2} =  - \left( {n + m} \right){{\vec a}_1} + \left( {n + 2m} \right){{\vec a}_2}
\end{array} \right.
\end{equation}
if gcd(n,3) = 1, and
\begin{equation}
\left\{ \begin{array}{l}
{{\vec t}_1} = \left( {\frac{n}{3} + m} \right){{\vec a}_1} + \frac{n}{3}{{\vec a}_2}\\
{{\vec t}_2} =  - \frac{n}{3}{{\vec a}_1} + \left( {\frac{2n}{3} + m} \right){{\vec a}_2}
\end{array} \right.
\end{equation}
if gcd(n,3) = 3 [gcd(\emph{p,q}) is the greatest common divisor of \emph{p} and \emph{q}]. Here, the vectors ${\vec a}_{1,2}$ are the primitive vectors of the bottom sheet as schematized in Fig. 1, i.e., ${\vec a_{1,2}} = {\vec r_{1,2}} - {\vec r_3}$ and their relationship with the primitive vectors ${\vec b_{1,2}}$ of the top sheet is
\begin{equation}
\left\{ \begin{array}{l}
{{\vec b}_1} = \left( {\cos {\phi _{TL}} - \frac{{\sin {\phi _{TL}}}}{{\sqrt 3 }}} \right){{\vec a}_1} + 2\frac{{\sin {\phi _{TL}}}}{{\sqrt 3 }}{{\vec a}_2}\\
{{\vec b}_2} =  - 2\frac{{\sin {\phi _{TL}}}}{{\sqrt 3 }}{{\vec a}_1} + \left( {\cos {\phi _{TL}} + \frac{{\sin {\phi _{TL}}}}{{\sqrt 3 }}} \right){{\vec a}_2}
\end{array} \right.
\end{equation}
The number of atoms in a primitive cell is $N_a = 4\left[ {{{\left( {n + m} \right)}^2} + m\left( {n + 2m} \right)} \right]$ in the former case and ${N_a} = 4\left[ {{m^2} + nm + {n^2}/3} \right]$ in the latter one.
To compute the electronic structure of these T-GBLs, we employed atomistic tight-binding calculations as in \cite{pere09b,lais10,moon12,sant12,hung14a,chun14}. A uniform uniaxial-strain of angle $\theta$ with respect to the Ox axis is applied in the in-plane direction (see Fig. 1). This strain causes changes in the $C-C$ bond vector $\vec r_{ij}$ according to ${\vec r_{ij}}\left( \sigma  \right) = \left\{ {\mathds{1} + {M_{strain}}\left( {\sigma ,\theta } \right)} \right\}{\vec r_{ij}}\left( 0 \right)$ with the strain tensor
\begin{equation}
  {M_{strain}} = \sigma \left[ {\begin{array}{*{20}{c}}
{{{\cos }^2}\theta  - \gamma {{\sin }^2}\theta }&{\left( {1 + \gamma } \right)\sin \theta \cos \theta }\\
{\left( {1 + \gamma } \right)\sin \theta \cos \theta }&{{{\sin }^2}\theta  - \gamma {{\cos }^2}\theta }
\end{array}} \right]
\end{equation}
where $\sigma$ represents the strain and $\gamma \simeq 0.165$ is the Poisson ratio \cite{blak70}. Taking into account the strain effects, the hopping parameters were adjusted accordingly as in \cite{pere09b}. Note that due to the lattice symmetry, the bandstructure with the applied strain of angles $\theta$ and $\theta + 60^\circ$ has the same properties and hence our investigation is limited to $\theta \in [0^\circ,60^\circ]$. Here, we especially focus on the possibility of achieving low energy saddle points. Hence, since the bandstructure of T-GBLs is nearly symmetrical around the neutrality (Dirac) point in the considered energy range \cite{shal10,moon12}, we only present and analyze the properties of conduction (positive-energy) bands throughout this work.

To analyze the basic properties of the bandstructure of T-GBLs under strain, we display pictures of the lowest energy bands of the lattice $\phi_{TL} = 9.43^\circ$ (i.e., \emph{n} = 1 and \emph{m} = 3) without and with strain of $\left(\sigma,\theta\right)=\left(6\%,20^\circ\right)$ in Fig. 2(a) and 2(b), respectively. It is well known that in unstrained graphene monolayers, the first Brillouin zone has a hexagonal form with six Dirac cones at their corners. Additionally, these corners are divided into two sets of inequivalent points, i.e., $K$ and $K'$. In unstrained T-GBLs, these Dirac points of two single layers are folded back to two Dirac points, $K$ and $K'$, in the reduced Brillouin zone \cite{moon12}. Therefore, the Brillouin zone of T-GBLs has the same hexagon form as that of single layers, however, their lowest band is composed of a pair of nearly degenerate branches around their Dirac (or \emph{K}) points. This is indeed illustrated in Figs. 2(a) and 2(c). Interestingly, we find here that when a strain is applied, (i) this degeneracy can be totally lifted (see Figs. 2(b) and 2(d)) and (ii) the Dirac cones of T-GBLs are no longer located at their \emph{K}-points. The latter is a general phenomenon observed for 2D graphene systems, which has been well explained by the effects of strain on the distance and hopping energies between \emph{C}-atoms. The former can be explained as a consequence of the different orientations of the two graphene layers in this lattice. Basically, as reported in \cite{chun14,kuma12}, the strain-induced deformation of the graphene bandstructure, i.e., resulting in the displacement of Dirac cones from the $K$-point, is strongly dependent on the lattice orientation. The two graphene layers in the T-GBLs have different orientations with respect to the strain direction and hence the strain effects on their bandstructure are different \cite{hung14b}. Though they are coupled in the T-GBLs, this difference still manifests itself on the separation of the degenerate bands around the Dirac cones and hence explains essentially our results. Here, we distinguish two groups of Dirac cones, $D$ and $D'$ ($D = 1-6$), which form two similar irregular hexagons and are just shifted from each other in the k-space. As a first important conclusion, our calculations demonstrate that the degeneracy of the bands around the Dirac cone of T-GBLs is broken and hence the number of Dirac cones in the first Brillouin zone of T-GBLs can double when a strain is applied. This result has not been observed yet in the previous studies based on the continuum approximation and the analysis of strain effects in the extended Brillouin zones, i.e., in the original Brillouin zones of single layers.

\begin{figure}[!t]
\centering
 \includegraphics[width=3.4in]{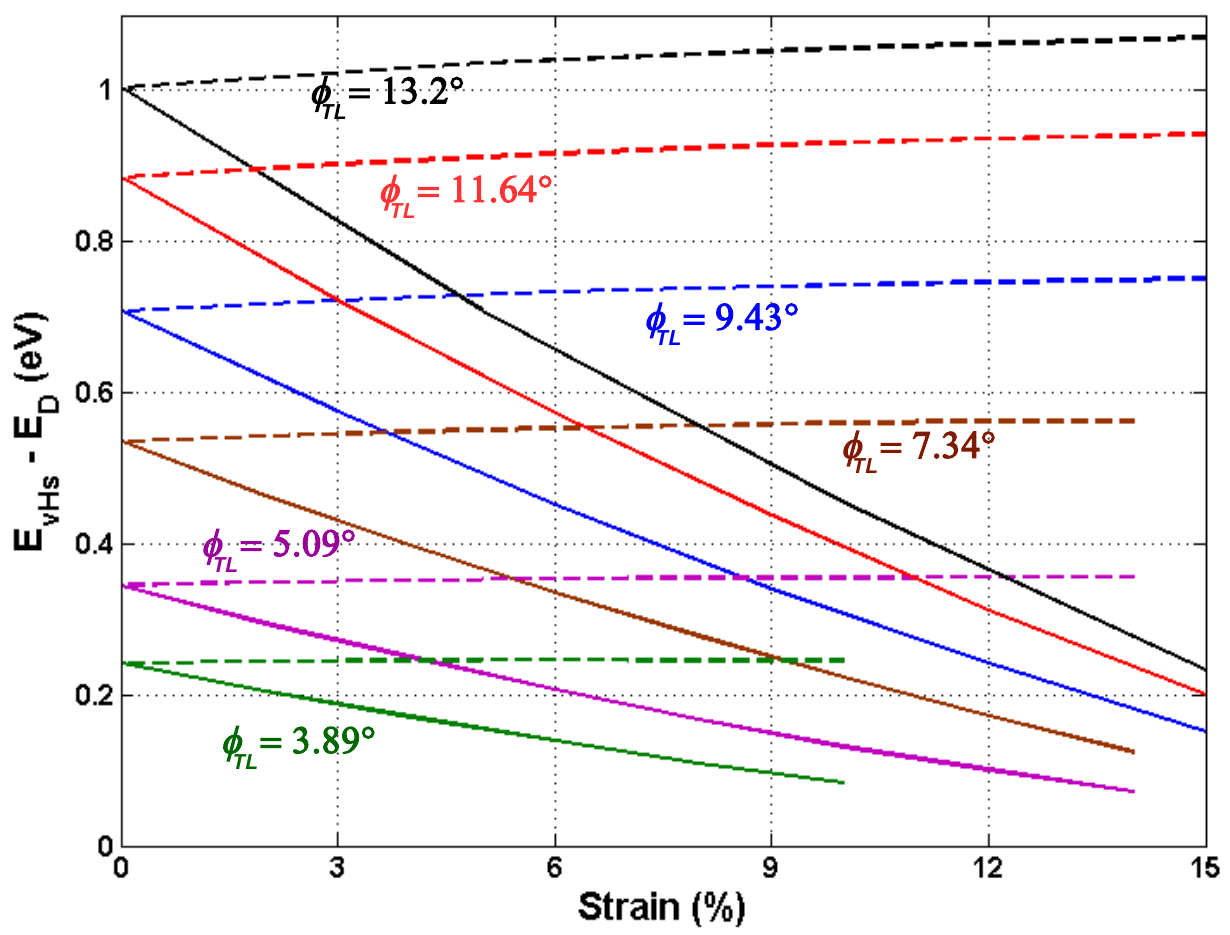}
\caption{Energy spacing between van Hove singularity and neutrality points as a function of strain strength for different twist angles $\phi_{TL}$. The strain direction is $\theta = 30^\circ$ in all cases except for the lattice with $\phi_{TL} = 11.64^\circ$, where $\theta = 0^\circ$. The solid and dashed lines correspond to the low and high energy saddle points, respectively.}
\label{fig_sim4}
\end{figure}

Next, we discuss the consequences of the phenomenon observed above on the properties of van Hove singularities. In Fig. 2(d), we see that the low energy bands joining the Dirac points $D$ and $D'$ of the considered lattice are actually crossing bands. Therefore, even if their relative shift can occur when a strain is applied, there is no saddle point between them. However, as shown in Fig. 3, saddle points occur close to the line joining Dirac cones of the same group. Here, we would like to notice that because of the interaction between the two layers, the bands in T-GBLs are more complex than that of single layers, especially, under strain. Actually, the saddle points occur close to but not exactly on the line joining Dirac cones in the strained lattices. As discussed later, this feature can result in difference of bandstructure properties between two different T-GBL types corresponding to gcd(n,3) = 1 and gcd(n,3) = 3. More important, we find as displayed in Fig. 3 that besides the separation of Dirac cones discussed above, the saddle points are also separated in energy, which is a simple consequence of the irregularity of hexagons connecting Dirac cones when the strain is applied. Note that in the unstrained lattices, these saddle points are identical, i.e., are found at the same energy (see in Fig. 3(a)). In strained lattices, it is thus possible to achieve three different saddle points (see in Fig. 3(b)). The existence of these saddle points is also confirmed by our plots of density of states (DOS) in the insets of Fig. 3. We also observe two other features. First, compared to the unstrained case, some of the saddle points of strained T-GBLs are located at lower energy while the others are formed at higher energy. Second, because of their separation, the peak of DOS at these saddle points is generally smaller than that of unstrained T-GBLs. Generally, our results agree well with the experiments \cite{myan13} and theoretical prediction based on the continuum approximation \cite{dchu13}, i.e., the strain can lower the energy of saddle points. However, the energy separation of saddle points and their properties mentioned above are new features observed here.
\begin{figure*}[!t]
\centering
\includegraphics[width=5.4in]{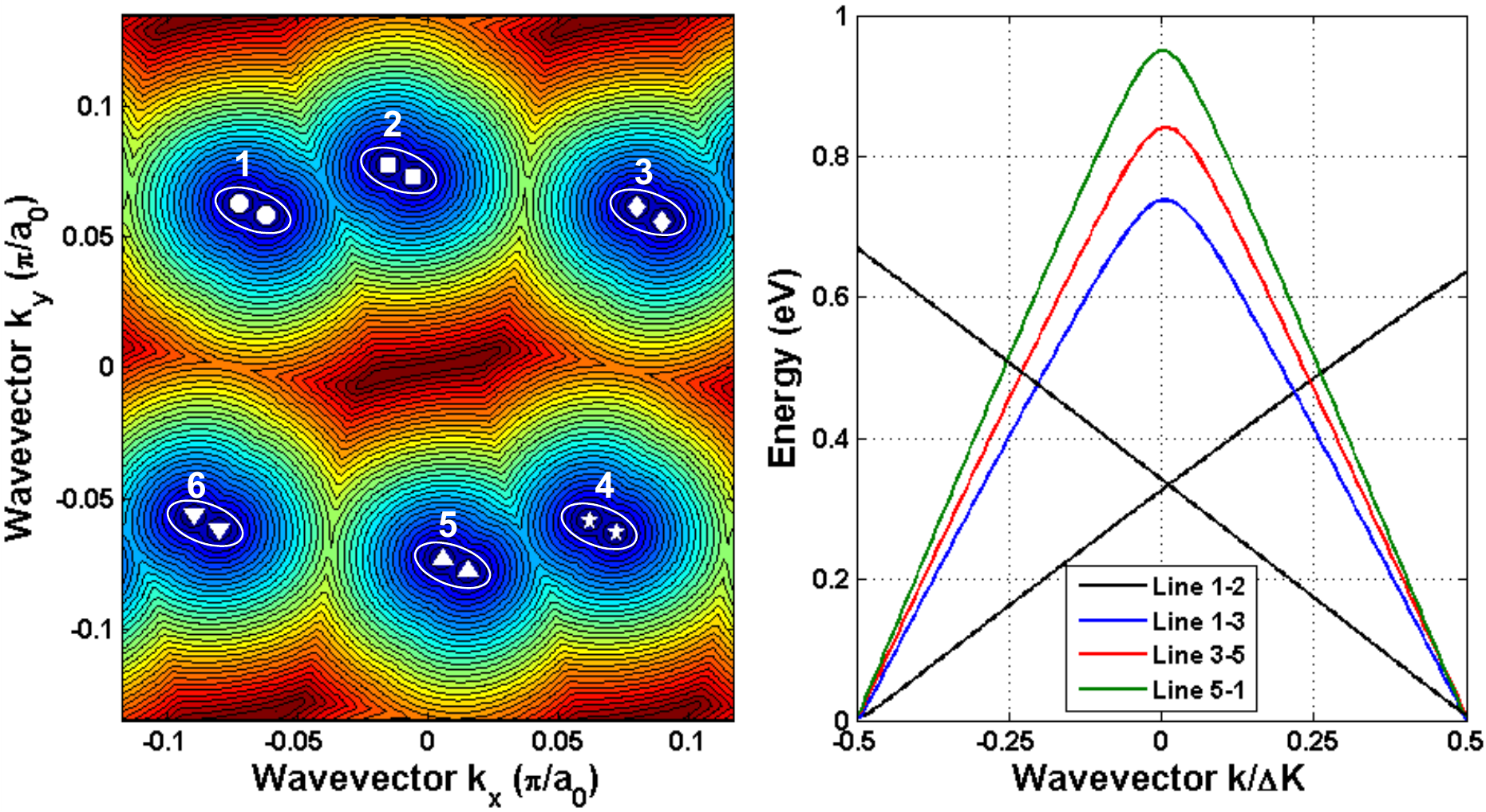}
\caption{Map of the lowest positive-energy bands (left) and bands joining Dirac points (right) in the lattice of twist angle $\phi_{TL} = 11.64^\circ$, i.e., with gcd(n,3) = 3. The strain $\left(\sigma,\theta\right)=\left(3\%,15^\circ\right)$ is applied.}
\label{fig_sim4}
\end{figure*}

All phenomena shown above, of course, are strongly dependent on the applied strain, i.e., on its strength and its applied direction. Now, we would like to discuss the properties of Dirac cones and saddle points with respect to the strain direction $\theta$. On the top of Fig. 4, we present two diagrams showing the position of Dirac cones ($D$ and $D'$) when changing $\theta$ from $-90^\circ$ to $90^\circ$. We additionally note that the effects of strain angles $\theta$ and $\theta + 180^\circ$ are identical. It is shown that when changing $\theta$, the Dirac points move around the K-points of unstrained lattices in well determined orbits. The form of hexagons $D$ and $D'$ connecting Dirac cones can be determined from the cone position in those orbits, according to the strain direction. Considering those orbits, we also confirm again that for the strain angles $\theta$ and $\theta+60^\circ$, the bands have the same properties, i.e., the hexagons in Figs. 4(a,b) for the strain angle $\theta$ is identical to that for angle $\theta+60^\circ$ after a rotation of $60^\circ$ and a translational displacement. On the bottom of Fig. 4, we plot the energy spacing between the saddle and neutrality (Dirac) points $\Delta E_{vHs} = E_{vHs} - E_D$ as a function of $\theta$ for both cases of tensile ($\sigma = 6\%$) and compressive ($\sigma = -6\%$) strains. We find that (i) in all strain cases, there is always at least one saddle point at lower energy than that of unstrained lattice and (ii) the saddle points have opposite properties for the tensile and compressive strains. In particular, the low energy saddle point is observed at the lowest energy for tensile strain while it is at the highest energy for compressive strain when $\theta = 30^\circ + i60^\circ$ and vice versa when $\theta = i60^\circ$. Additionally, two degenerate saddle points are achieved when $\theta = i30^\circ$. When comparing the two cases of tensile and compressive strains, it is shown that the saddle points are generally observed at higher energy in the compressive case than that in the tensile one. This can be explained by the fact that besides the deformation of the bandstructure, the hopping energies increase when the compressive strain is applied and hence the energy scale of all the bands is higher than that in the tensile case.

In the previous work \cite{dchu13}, Chu et al. have also investigated the effects of strain on the properties of saddle points of T-GBLs using the continuum approximation. They concluded that the tensile strain applied along the zigzag direction of the top layer (and compressive strain along the armchair direction) can lower the energy position of saddle point but the tensile strain along the armchair direction (and compressive strain along the zigzag direction) can increase it. We would like to notice that though our calculations show that the strain always lower the lowest energy saddle points, the data in \cite{dchu13} are qualitatively consistent with our results obtained for the second lines in Fig. 4(c) (determined at $\theta = 0$ from the bottom) for both strain types. The zigzag and armchair directions of top layer are actually $\theta = \phi_{TL}/2$ $(\simeq 4.72^\circ)$ and $\theta = 30^\circ + \phi_{TL}/2$ $(\simeq 34.72^\circ)$, respectively. However, the full properties of saddle points are shown to be more complex than what was reported in \cite{dchu13}.

Next, we discuss the possibilities of achieving low energy saddle points in T-GBLs by applying strain. In Fig. 5, we plot $\Delta E_{vHs}$ as a function of strain in different twisted lattices. All considered lattices are with gcd(n,3) = 1, except $\phi_{TL} = 11.64^\circ$ that is for gcd(n,3) = 3. To seek for the lowest energy saddle points, the strain direction $\theta = 30^\circ$ is considered in all cases except $\theta = 0^\circ$ for $\phi_{TL} = 11.64^\circ$. The differences between the two lattice types with gcd(n,3) = 1 and gcd(n,3) = 3 will be discussed below. Note that for these strain angles, there are only two saddle points in the considered energy range but they can separated in three points for other strain directions (see Fig. 4(c)). In the unstrained lattices, the data in Fig. 5 confirm a good agreement with what was observed in experiments, i.e., $\Delta E_{vHs}$ is almost linearly proportional to the twist angle $\phi_{TL}$ \cite{guli10,brih12}. When the strain is applied, while $\Delta E_{vHs}$ for high energy saddle point slightly increases, the $\Delta E_{vHs}$ for low energy saddle point significantly decreases with strain. Interestingly, we find that the larger the twist angle, the stronger the reduction of $\Delta E_{vHs}$ when increasing the strain. We suggest that this can be a consequence of the folding of the bandstructure from the original bands of single layers, i.e., the band folding can weaken the effects of bandstructure deformation in slightly twisted lattices. More important, our obtained results demonstrate the possibilities of tuning the position of saddle points by strain \cite{myan13} and of achieving low energy saddle points for a large range of $\phi_{TL}$, at the expense, of course, of lower peaks of DOS as discussed above. Moreover, it also suggests that multi ($>$ 1) saddle points can be obtained for moderate $\phi_{TL}$ and strain.
\begin{figure*}[!t]
\centering
\includegraphics[width=6.3in]{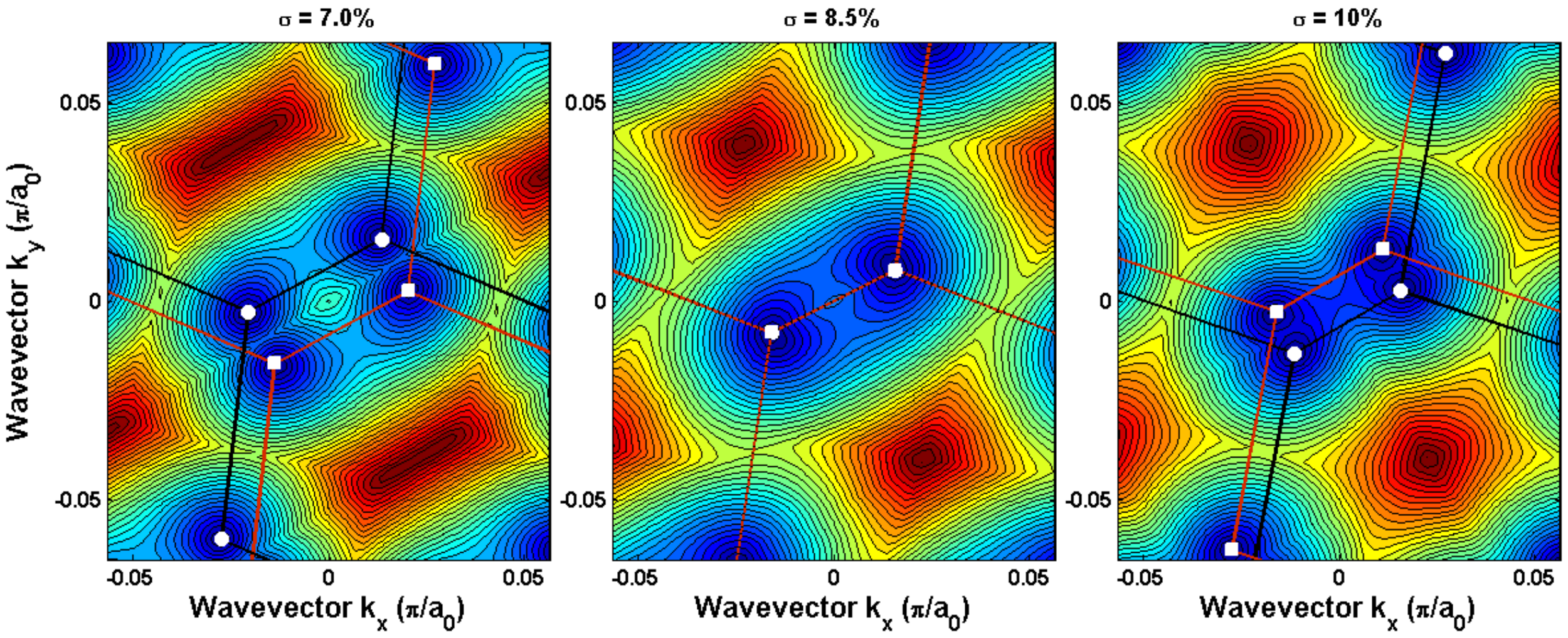}
\caption{Map of the lowest positive-energy bands for different strains in the lattice of twist angle $\phi_{TL} = 3.89^\circ$. The strain direction is $\theta = 30^\circ$.}
\label{fig_sim4}
\end{figure*}

Now, we would like to clarify the properties of the T-GBLs with gcd(n,3) = 3 detected from our calculations, which are different from the case of gcd(n,3) = 1. In Fig. 6, we present the cartography of the lowest energy bands (left) and the bands joining Dirac points (right) for the case of \emph{n} = 3 and \emph{m} = 7, i.e., $\phi_{TL} = 11.64^\circ$ in Fig. 5. Besides the phenomena discussed above, our calculations also show that the bandstructure of T-GBLs with gcd(n,3) = 3 has dramatically different properties, compared to the cases of gcd(n,3) = 1. In particular, instead of the existence of a saddle point as in the case of gcd(n,3) = 1, a crossing of bands is observed close to the line between Dirac points 1 and 2. The saddle points however occur close to the lines joining the Dirac points 1-3, 3-5 and 5-1. In addition, the $\theta$-dependence of $\Delta E_{vHs}$ in the two lattice types are also opposite, i.e., a tensile strain in the case of gcd(n,3) = 3 (not shown) has similar effects as the compressive one for gcd(n,3) = 1 shown in Fig. 4(c) and vice versa. We suggest that these differences are a direct consequence of the difference in lattice symmetries. Because of this difference, the interaction between two layers have different effects on the bandstructure when they are coupled in T-GBLs. In spite of these differences, the behavior of the VHS energy $\Delta E_{vHs}$ when increasing the strain is very similar in both lattice types, as observed in Fig. 5. The same feature for transport gap in the vertical devices made of two twisted graphene layers has been also observed \cite{hung14b}.

Finally, we further investigate the bandstructure when a large strain is applied. We find, as illustrated in Fig. 7 for the lattice of twist angle $\phi_{TL} = 3.89^\circ$, another interesting feature that the separation of Dirac cones in the cases of small strain can disappear when increasing the strain strength (fro $\sigma = 8.5\%$ in this case). By further enhancing the strain, this separation occurs again. This is actually a consequence of the fact that while the strain tends to displace the Dirac cones in the k-space, the size of the Brillouin zone in T-GBLs (especially, in the case of slightly twisted lattices) is much smaller than that of a single layer. Hence, the merging of Dirac cones occurs at large strain when their separation reaches the size of the Brillouin zone. The value of strain for which this merging occurs is, of course, dependent on the size of the Brillouin zone and hence on the twist angle, i.e., it increases when increasing $\phi_{TL}$. This property may have an important impact on the physical phenomena related to the Dirac fermions, i.e., to the properties of Dirac cones. As an example, it should have a strong impact on the transport gap in vertical devices made of twisted graphene layers \cite{hung14b} since this gap is essentially governed by the separation of Dirac cones of the two layers in the \emph{k}-space.

In conclusion, we have investigated the effects of uniaxial strain on the bandstructure of twisted graphene bilayer using atomistic tight-binding calculations including the detailed arrangement of \emph{C}-atoms. Compared to the previous studies based on the continuum approximation, our calculations show some new properties. In particular, the band degeneracy around the Dirac cones observed in unstrained lattices can be totally broken by strain while the bandstructure is dramatically deformed. It makes the number of Dirac cones in the first Brillouin zone doubles and results in the energy separation of van Hove singularity points. It is also shown that these phenomena are strongly dependent on the strength of strain, its applied direction, and the twist angle. Actually, the van Hove singularity points can be efficiently modulated and hence the possibility of observing this phenomenon at low energy is demonstrated in a large range of twist angle (i.e., larger than $10^\circ$). Hence, our results provide good guidelines for exploiting the strain effects to modulate the electronic phenomena related to the properties of Dirac cones and the existence of van Hove singularities at low energy in this type of graphene lattice.

\textbf{\textit{Acknowledgment.}} This research in Hanoi is funded by Vietnam's National Foundation for Science and Technology Development (NAFOSTED) under grant number 103.01-2014.24.

\end{document}